\documentclass[twocolumn,aps,pra,showpacs]{revtex4}

\usepackage[dvips]{graphicx}
\usepackage{amsmath}
\usepackage{SIunits}
\usepackage{ulem}

\begin{document}

\title{Measurement of the atom number distribution in an optical tweezer using single~photon counting}
\author{A. Fuhrmanek, Y.R.P. Sortais\footnote{corresponding author: yvan.sortais@institutoptique.fr}, P. Grangier, A. Browaeys}

\affiliation{Laboratoire Charles Fabry, Institut d'Optique, CNRS, Univ Paris-Sud,
Campus Polytechnique, 2 avenue Augustin Fresnel, RD 128,
91127 Palaiseau cedex, France}

\date{\today}

\begin{abstract}
We demonstrate in this paper a method to reconstruct the atom number distribution of a cloud containing a few tens of cold atoms. The atoms are first loaded from a magneto-optical trap into a microscopic optical dipole trap and then released in a resonant light probe where they undergo a Brownian motion and scatter photons. We count the number of photon events detected on an image intensifier. Using the response of our detection system to a single atom as a calibration, we extract the atom number distribution when the trap is loaded with more than one atom. The atom number distribution is found to be compatible with a Poisson distribution.
\end{abstract}

\pacs{37.10.-x, 05.40.Jc}

\maketitle

\section{Introduction}
The last few years have seen a growing interest in the study of mesoscopic systems consisting of typically a few tens of interacting particles. The properties of these systems usually cannot be described by a mean-field approach and are already too complicated to be calculated from the behavior of each individual interacting particle. Dense clouds of ultra-cold atoms  provide an ideal test bed to study these mesoscopic systems. The interactions are well-understood at the two-body level and, experimentally, one benefits from the host of tools developed over the years to investigate ultra-cold atomic clouds (see e.g.~\cite{BlochRMP08}). Various implementations have already been realized such as arrays of optical tweezers~\cite{Dumke02} or of magnetic traps~\cite{Whitlock09}, double-well potential geometries~\cite{Albiez05} and optical lattices~(e.g.\cite{Mandel03,Nelson2007}).

Beyond their intrinsic interest, a possible application of dense clouds of ultra-cold atoms  is the production of atomic states for which the fluctuations of the number of atoms are reduced with respect to the Poissonian case. These clouds could for example be useful for atomic interferometry below the standard quantum limit~\cite{Wineland94}. Recent demonstrations have been achieved using Bose-Einstein condensates with a few thousand atoms (\cite{Orzel01,Esteve08}). The clouds could also be used as a source delivering a given number of atoms, as was already demonstrated in the single atom case in an optical tweezer~\cite{Schlosser2001} or in a one-dimensional optical lattice~\cite{Forster06}. For larger numbers of atoms, experiments showing the reduction of the atom-number fluctuations have been reported in Bose-Einstein condensates in an optical trap~\cite{Chuu2005}, in optical lattices~\cite{Itah10}, or in arrays of magnetic micro-traps~\cite{Whitlock10}. For all these applications the knowledge of the distribution of the number of atoms, and not only the average number of atoms, is essential.

The number of atoms in a cold atomic cloud can be measured by usual optical techniques, such as laser induced fluorescence or absorption. Nevertheless, special care must be taken when the number of interacting particles involved is small. For instance, when using fluorescence imaging
the small number of atoms combined with the low collection efficiency makes it hard to collect more than a few photons per realization of the experiment, therefore preventing the reliable extraction of the number of atoms in single shot. To circumvent this problem,
a method was recently demonstrated~\cite{Bucker09} where freely propagating atoms fall through a sheet of resonant light leading to the detection of many photons per atom.

In this  paper we present a method to measure the distribution of the number of atoms trapped in a microscopic optical dipole trap in a regime where the number of atoms is a few tens. The method relies on single-photon counting. The principle is as follows: we release the cloud of cold atoms from the dipole trap and let them expand in a resonant light probe. The atoms diffuse in the probe in a Brownian motion and emit photons. Some of the photons are collected by an imaging system and impinge on an image intensifier where we count them. By repeating the experiment several times and recording the counting results, we reconstruct the distribution of the number of detected photons. Our experimental setup can work either with exactly one atom, or with a cloud of up to a few tens of atoms. In this way we  perform a calibration of the number of detected photon events when a single atom is trapped, and use this calibration to determine the atom number distribution.

The paper is organized as follows. In Sect.~\ref{section:Setup} we describe briefly the setup. Section~\ref{section:Loadingdipoletrap} describes the loading of our microscopic dipole trap in the single-atom regime and in the many-atom regime. In section~\ref{section:expprocedure} we describe the experimental sequence and in Sect.~\ref{section:image_analysis} we present the image analysis used to count the number of photon events. Sect.~\ref{section:SingleAtom} deals with the calibration of our detection system using a single atom. Section~\ref{section:multiatom} explains how we extract the atom number distribution in the multi-atom regime. In Sect.~\ref{section:counter_test} we compare the mean number of atoms extracted by this counting method to the result obtained by direct integration of the laser induced fluorescence, and find good agreement between the two methods.

\section{Experimental setup}\label{section:Setup}
Our experimental setup has been described in details elsewhere~\cite{Sortais07,Tuchendler08,Fuhrmanek2010} and consists essentially in a microscopic optical dipole trap, in which we trap up to a few tens of $^{87}$Rb atoms  (see details in Section~\ref{section:Loadingdipoletrap}). The trap is produced by a laser beam at $850$\,nm that is focused by an aspheric lens with a numerical aperture $\textrm{N.A.} = 0.5$. The resulting spot has a Gaussian profile with a waist $w~=~1.1~\mu$m. A power of $9$\,mW yields a trap depth of $2$\,mK. The dipole trap is loaded from a cold atomic cloud held in a magneto-optical trap produced around the focal point of the aspheric lens. The magneto-optical trap captures atoms from a Zeeman decelerated atomic beam coming from
an oven.

We use the same aspheric lens to collect the fluorescence light at $780$\,nm. A dichroic beam-splitter separates the fluorescence light from the dipole trap light. A fraction of the fluorescence light is sent to a fiber-coupled avalanche photodiode operating in a single photon counting mode. The remaining fluorescence is  directed onto an image intensifier, followed by a low-noise CCD camera. We have measured a detection efficiency $\eta \simeq 3\times 10^{-3}$, which takes into account the solid angle of the
aspheric lens ($7\%$), the transmission of the optics ($40\%$), and the measured quantum efficiency of the intensifier photocathode ($10\%$).
Here, we note that using an image intensifier is crucial for the experiment presented in this paper because it enables our imaging system to detect single photon events. In the absence of the intensifier, single photon events have an amplitude lower than the noise of the CCD camera alone ($6\rm e^-$/pixel). The intensifier amplifies a single photon event to an average amplitude of $920\rm e^-$/pixel, well above this noise level.

In order to characterize the atomic cloud, we use a probe consisting in two counter-propagating  laser beams in a $\sigma_{+} - \sigma_{-}$ configuration. This probe is tuned to the $(5^{2}S_{1/2}, F=2)$ to $(5^{2}P_{3/2}, F'=3)$ transition and is sent together with a repumping light tuned to the $(5^{2}S_{1/2}, F=1)$ to  $(5^{2}P_{3/2}, F'=2)$ transition. The probe light is focused down to a waist of $0.9$\,mm and has a saturation parameter $I/I_{\rm sat} \approx 2.7$ for each beam.

\section{Loading of the dipole trap}\label{section:Loadingdipoletrap}
The experiment can operate either in a regime where only one atom is present in the dipole trap or in a regime where up to $\sim10$ atoms are trapped. The transition between the two regimes is governed by the atomic density in the magneto-optical trap, which is controlled by the flux of the atomic beam. The average number of atoms $\bar{N}$ in the dipole trap is set by two competing processes: the loading from the magneto-optical trap, and the two-body losses due to collisions assisted by the near-resonant light of the cooling and trapping lasers~\cite{Schlosser2002}. When the loading rate of the dipole trap is lower than the light-assisted collision rate, only one atom can be present at a time in the trap: if a second atom enters the trap a light-assisted collision takes place and both atoms are lost. On the contrary, when the loading rate of the dipole trap overcomes the inelastic loss rate, the average number of atoms is larger than one~\cite{Schlosser2002,Kuppens2000}. For the experiments presented below we find that, after a fixed loading time ($\lesssim1$\,s typically), the mean number of atoms in our microscopic trap increases linearly with the trap depth and reaches $\bar{N}=9$ for a trap depth of $\sim20$\,mK. We measure the temperature of the trapped atom(s) using either a release-and-recapture method~\cite{Tuchendler08} or  a time-of-flight method~\cite{Fuhrmanek2010}. We find a temperature around $150$~$\mu$K in the single atom case and around 1~mK in the multi-atom regime.

In the single-atom regime, the presence of an atom in the trap is revealed by a step in the fluorescence signal collected on the APD~\cite{Sortais07}. This step is used to trigger an imaging sequence where the number of fluorescing atoms is exactly $1$: this way there are no fluctuations in the atom number from shot to shot (variance $\Delta N^2 = 0$). In the multi-atom regime, the situation is different as we observe no such steps on the APD signal: firstly because when the atomic density increases, the time spent on average by each atom in the trap decreases below the time bin resolution of the APD counter due to an increasing inelastic loss rate, and secondly because the height of the steps itself decreases due to light-induced dipole-dipole interactions that shift and broaden the  atomic line~\cite{Morice95}. One thus cannot rely on fluorescence steps to perform an \textit{in situ} measurement of the number of trapped atoms. Instead, the technique presented in this paper goes around this problem by converting the multi-atom signal into a sum of single atom events, spatially separated on a CCD camera. In this second regime, the absence of fluorescence steps on the APD signal also makes it impossible to trigger the imaging sequence after a given number of atoms is loaded in the trap. We thus trigger the imaging sequence after a fixed loading duration. Due to the random nature of the two-body losses during the loading this procedure leads to
fluctuations in the number of trapped atoms  from shot to shot ($\Delta N^2 \neq 0$)~\cite{Schlosser2002}.

\section{Experimental procedure and fluorescence images}\label{section:expprocedure}
Once the trap is loaded, either with exactly one atom, or with $N$ atoms, we switch off the cooling and repumping lasers and wait an additional $30$\,ms before switching off the dipole trap. This waiting period  allows enough time for the atoms of the magneto-optical trap to leave the observation region, otherwise they would strongly contribute to the signal observed with the intensifier. We then switch off the dipole trap (in $\sim 200$\,ns) and, after a time of flight of the atoms of $1~\mu$s in free space, we turn on the resonant probe, while gating the intensifier for a given period of time $\Delta t$ (in the $10-60~\mu$sec range). Finally, we read out the CCD camera, which lasts $\sim500$\,ms, while launching the next loading sequence.

During the $1~\mu$s time of flight in free space, the atoms start separating from each other. Then they walk randomly in the retro-reflected light probe and scatter photons at a rate $\sim \Gamma/2\simeq 2\times 10^7$\,s$^{-1}$ ($\Gamma$ is the line-width of the atomic transition). Some of the scattered photons are collected on the image intensifier  and impinge on the CCD at different positions due to the random walk of the atom in the probe light~\footnote{The detected photon events are distributed around the geometrical images of the atoms according to the point spread function of the imaging system, which includes diffraction, residual aberrations and defocus due to the spatial distribution of the atoms. Given the size of the tweezer and the temperature of the atoms, the aberrations and the defocus are negligible. The collection efficiency therefore does not depend on the position of the atoms.}. The total number $n$ of detected events is  proportional to the number of atoms $N$ on average.

\begin{figure}
\includegraphics[width=8cm]{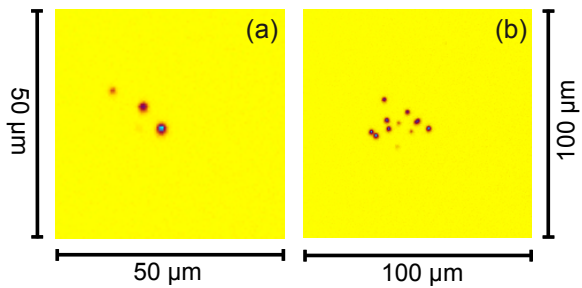}
\caption{(Color Online) Figure (a) is an image taken for a $40\mu$s-illumination period of a single atom. The region of interest has a size of $50~\mu$m$\times 50~\mu$m in the plane of the atoms, corresponding to a measured temperature of $150~\mu$K (see text). In figure (b), about $8$ atoms were released from the trap and illuminated with a $20~\mu$s probe pulse. The image size is $100~\mu$m$\times 100~\mu$m in the plane of the atoms, corresponding to a temperature of $1$\,mK.}
\label{figure:PhotonEvent_on_CCD}
\end{figure}

The images shown in Fig.~\ref{figure:PhotonEvent_on_CCD}(a) and (b) were obtained after a single realization of the experimental sequence. Figure~\ref{figure:PhotonEvent_on_CCD}(a) corresponds to the time of flight of a single atom moving randomly in the probe light. Figure~\ref{figure:PhotonEvent_on_CCD}(b) corresponds to the time of flight of about $8$ initially trapped atoms. In both cases, the detected photons correspond in a large extent to photons being scattered by the atom and detected by the intensifier and, in a smaller extent,  to photons due to spurious laser light or self-induced charges generated inside the intensifier (see more details below). This fact led us to adjust the size of the region of interest to the measured temperature of the atom(s): for a time of flight of duration $\Delta t$, we chose a square with a side dimension $\sim 4\sigma_{v}\Delta t$, where $\sigma_{v}=\sqrt{k_{\rm B} T/M}$, $T$ being the temperature of the atoms, $M$ their mass, and $k_{\rm B}$ the Boltzman constant. This choice leads to a negligible probability of missing an atomic event, while reducing the number of background events, which are uniformly distributed on the CCD.

\section{Procedure for analyzing the images}\label{section:image_analysis}
In order to count the number of events detected on the CCD camera, one has to decide what one calls an event. In order to do so, we first characterize the noise of the CCD camera, which is essentially due to the readout of the CCD chip. We find a root-mean-square (rms) noise $\sigma_{\textrm{CCD}} = 6{\rm e}^-$/pixel. We then apply the following procedure. We search for the pixel with the largest amplitude of the signal. If the signal amplitude on this pixel is larger than $6\sigma_{\textrm{CCD}}$, we call it a peak and exclude from the image a zone of $13\times 13$ pixels around it, the rms size of an individual event being $1.3\times 1.3$\,pixels (set by the intensifier resolution and corresponding to $0.6\times0.6~\mu$m$^2$ in the plane of the atoms)~\footnote{The exclusion process may lead to missing neighboring peaks in the $13\times 13$ pixels zone around the main peak. We have therefore developed a specific routine to treat these occurrences, which represent $\sim10\%$ of the cases.}. We then repeat this procedure by searching for the pixel with the next largest amplitude until no peak with an amplitude larger than $6\sigma_{\textrm{CCD}}$ is found.

We have independently measured the amplitude distribution of single-photon events recorded by the intensifier followed by the CCD camera. Using this distribution, we have found that less than $4\%$ of the events have a peak amplitude smaller than $6\sigma_{\textrm{CCD}}$. Our procedure for counting the events therefore does not underestimate their number by more than $4\%$.

\section{Calibration  using a single atom}\label{section:SingleAtom}
\begin{figure}
\includegraphics[width=8cm]{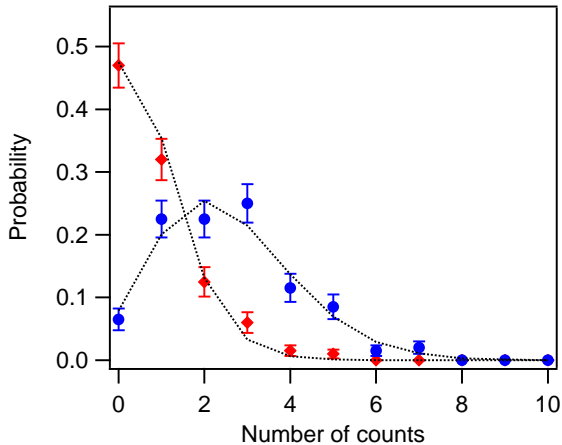}
\caption{(Color Online) Probability distribution of the number of detected events in one image. The probe duration is $30~\mu$s. Red diamonds : background events (no atom present in the trap). Blue circles : total number of detected events (including background events) when the dipole trap is filled with exactly one atom. In both cases, the probability $p$ to measure a given number a counts in one image is deduced by analyzing $200$ images. The error bar on each data point is calculated by $\sqrt{p(1-p)/200}$, which we tested by repeatability measurements. A Poisson distribution fits the data very well in both cases (dotted lines).}
\label{figure:SingleAtom_PhotonDistr}
\end{figure}

The first step of the calibration of the imaging system as a counting device was to measure the histogram of the number of background events in one image when no atom is trapped in the dipole trap. The background events have three different origins: the largest contribution comes from light scattering by the atoms of the atomic beam that cross the region of observation ($60\%$); the second largest contribution comes from  self-induced events generated by the gated intensifier when the high voltage is switched on and off ($33\%$ of the counts); finally, a smaller contribution comes from scattering of the probe beams on the surfaces inside the vacuum chamber ($7\%$). The histogram of the total number of background events is shown in Fig.~\ref{figure:SingleAtom_PhotonDistr}, which results from the analysis of $200$ images of the background. The probe pulse duration was $\Delta t=30~\mu$s. The data are well fitted by a Poisson distribution with average $\bar{ n}_{\rm bg}=0.7$.

In a second step, we repeated the same experiment with exactly one atom trapped in the dipole trap. In particular, the probe duration was the same as above ($\Delta t=30~\mu$s). As the detection efficiency is small and corresponds to a random collection of photons, we also expect a Poisson distribution for the number of detected photons emitted by the atom. In this case, the distribution of detected events, including the background events, is a Poisson distribution with a mean value:
\begin{equation}\label{equation:average_number_of_detected_events}
\bar{n}_{\rm single} = \bar{ n}_{\rm at} + \bar{n}_{\rm bg}\ .
\end{equation}
The experimental distribution is shown in Fig.~\ref{figure:SingleAtom_PhotonDistr}, together with a fit by a Poisson distribution. We obtain a good agreement for $\bar{n}_{\rm single}=2.5$. This yields an average number of detected photons, emitted by one atom, of $\bar{n}_{\rm at}=1.8$.

\begin{figure}
\includegraphics[width=8cm]{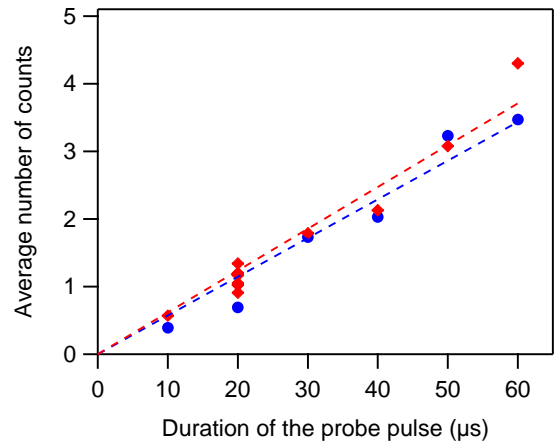}
\caption{(Color Online) Average number of detected photons after scattering by a single atom with temperature $16~\mu$K (blue circles) or $150~\mu$K (red diamonds), versus the duration of the atom illumination by the probe light. Linear fits to the data (dashed lines) yield  a photon detection rate $R_{\rm d}=57200\pm3100~\textrm{s}^{-1}$ for $16~\mu$K and $R_{\rm d}=61800\pm2400$\,s$^{-1}$ for $150~\mu$K.}
\label{figure:SingleAtom_Calibration}
\end{figure}

Finally, we tested the linearity of our counting system with the duration of the probe pulse. The results are shown in Fig.~\ref{figure:SingleAtom_Calibration}, where we plot the average number of detected events due to scattering by a single atom, $\bar{n}_{\rm at}$, obtained after subtraction of the average number of background events. We first performed the experiment with a single atom with a temperature $T\sim150~\mu$K and observed a linear dependency of $\bar{n}_{\rm at}$ with the duration of the probe illumination. A linear fit to the data yields a photon detection rate, for one atom released from the trap, of $R_{\rm d}~=~57200\pm3100~{\rm s}^{-1}$ (the error bar is from the fit). Knowing the detection efficiency of the imaging system ($\sim 3\times10^{-3}$), we deduce a scattering rate of $1.9\times 10^7~{\rm s}^{-1}$, in  agreement with the value calculated for the parameters of our probe.

In order to further test the linearity of our counting system, we also performed the same measurement for an atom adiabatically cooled down to $16~\mu$K~\cite{Tuchendler08}. In this case, the detected events tend to accumulate on a smaller area on the I-CCD, due to the smaller velocity of the atom (the time of flight of the atom is maintained constant and equal to $1~\mu$s with respect to the previous case at $T\sim150~\mu$K). In spite of the increase in the surface density of detected events, we were still able to discriminate between individual events and found the same linear dependency as above, with a photon detection rate of $61800\pm2400$\,s$^{-1}$.

To conclude, we detect on average $\bar{n}_{\rm at}=1.1$ photons scattered by a single atom for a probe duration $\Delta t=20~\mu$s. This value will serve as a calibration in the experiments described below, where the average number of atoms is the unknown.

\section{Atom number distribution in the multi-atom regime}\label{section:multiatom}
\begin{figure}
\includegraphics[width=8cm]{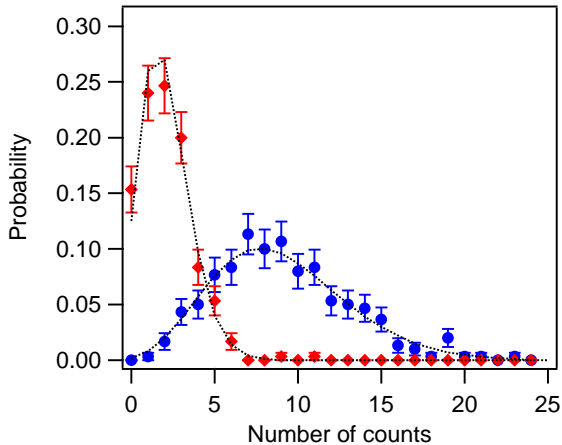}
\caption{(Color Online) Probability distribution of the number of detected events when the dipole trap is filled with $6.3$ atoms on average (blue circles). The data are well-fitted by a composed law that includes a Poisson law for the number of atoms in the trap. The data in red diamonds correspond to the background events only ($\bar{n}_{\textrm{bg}}=2.1$) fitted by a Poisson law. The duration of the probe is set to $20~\mu$s so that the average number of detected photons scattered per atom is $\bar{n}_{\textrm{at}}=1.1$.}
\label{figure:MultiAtom_PoissonComposed}
\end{figure}

We now operate in the regime where more than one atom are loaded in the dipole trap. We repeat the same experimental procedure and extract the histogram of the number of detected events for  a probe duration $\Delta t=20~\mu$s. Figure~\ref{figure:MultiAtom_PoissonComposed} shows the histogram of the background events as well as the probability distribution of the total number of detected events (including the background events) when about $6.3$ atoms are loaded in the dipole trap on average. The background gives an average number of events $\bar{n}_{\rm bg}=2.1$, larger than in Sect.~\ref{section:SingleAtom}, as we needed to increase the flux of the atomic beam in order to load more than one atom in the trap.

To fit the distribution of events in the multi-atom regime, we consider the distribution of the number of atoms $N$ in the dipole trap, $P(N,\bar{N})$, with  $\bar{N}$ the mean atom number. The probability to detect $n$ photon events is then given by the composed law~\cite{Schlosser2002}:
\begin{equation}\label{equation:composedlaw}
p(n) = \sum_{N=0}^{\infty} P(N,\bar{N})\times\Pi(n,N\times \bar{n}_{\rm at}+\bar{n}_{\rm bg})
\end{equation}
with $\Pi(n,\alpha)$ the Poisson distribution of mean $\alpha$. Even in the case when $P(N,\bar{N})$ is a Poisson law, the distribution $p(n)$ is not Poissonian. Nevertheless, for any distribution $P(N,\bar{N})$, the mean value of this composed law is given by:
\begin{equation}\label{equation:average_number_of_detected_events_multi}
\bar{n}_{\textrm{multi}}=\bar{N}\times \bar{n}_{\rm at}+\bar{n}_{\rm bg}\ .
\end{equation}
A direct calculation of the mean of the data shown in Fig.~\ref{figure:MultiAtom_PoissonComposed} yields $\bar{n}_{\textrm{multi}}=9$. Taking into account that the probe duration has been chosen to detect $\bar{n}_{at}=1.1$ events per atom and that $\bar{n}_{\textrm{bg}}=2.1$, equation~(\ref{equation:average_number_of_detected_events_multi}) yields an average number of atoms $\bar{N}=6.3$. Taking a Poisson distribution for $P(N,\bar{N})$, a fit of the data by the composed law~(\ref{equation:composedlaw}) leads to the same result. The result of the fit is shown in Fig~\ref{figure:MultiAtom_PoissonComposed}.

We now discuss the Poissonian assumption mentioned above for the atom number distribution $P(N,\bar{N})$. We extract from the data the variance of the number of detected events, $\Delta n_{\textrm{multi}}^2=\overline{n_{\textrm{multi}}^2}-\bar{n}_{\textrm{multi}}^2$. This variance is related to the variance of the number of atoms $\Delta N^2$ by the following expression, calculated using the probability $p(n)$ of equation~(\ref{equation:composedlaw}), and valid for any distribution $P(N,\bar{ N})$:
\begin{equation}\label{equation:variance_number_of_detected_events_multi}
\Delta n_{\textrm{multi}}^2=\bar{n}_{\rm at}^2\times\Delta N^2+ \bar{n}_{\textrm{multi}}\ .
\end{equation}
Taking again $\bar{n}_{\textrm{at}}=1.1$, we find for the data of Fig.~\ref{figure:MultiAtom_PoissonComposed} a
ratio of $\Delta N^2/\bar{N}~=~0.86\pm0.13$ (the error bar is statistical). This value is compatible with $P(N,\bar{N})$ being a Poisson distribution. We repeat this measurement for various average numbers of atoms ranging from $\bar{N}=2$ to $\bar{N}=9$ and find that $\Delta N^2/\bar{N}$ is equal to $0.76$ in average with a rms dispersion of $0.13$. This $1~\sigma$ uncertainty does not clearly indicate a sub-Poissonian behavior as $\Delta N^2/\bar{N}$ does not deviate from $1$ by more than $2~\sigma$.

\section{Comparison with fluorescence integration}\label{section:counter_test}
We have checked that the average number of atoms extracted by the counting method described so far is consistent with the number of atoms extracted by a second method based on fluorescence integration. This second method consists in illuminating the cloud of freely propagating atoms by a $2~\mu$s probe pulse, and accumulating the fluorescence detected on the CCD over many realizations of the experiment. In each time-of-flight experiment, we let the atoms fly during $1~\mu$s before we illuminate them with the probe pulse. As explained in \cite{Fuhrmanek2010}, after typically several hundreds repetitions of the experiment, the image shows a nearly gaussian distribution, reconstructed from many individual detected events. We perform this experiment in the multi-atom regime and then in the single-atom regime. We use the latter to extract the number of atoms in the multi-atom regime, by calculating the ratio of the integrated fluorescence values obtained in the two regimes.

\begin{figure}
\includegraphics[width=7cm]{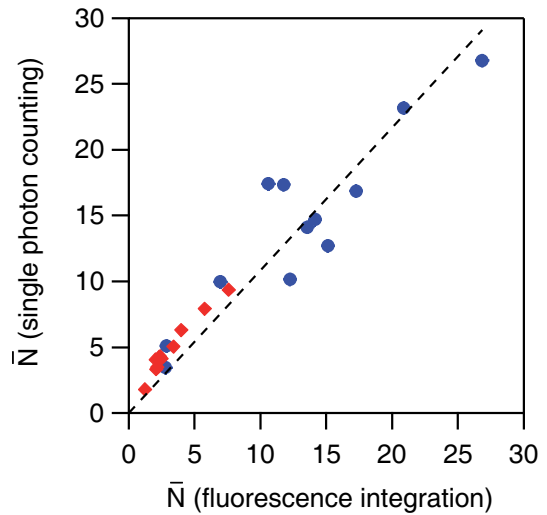}
\caption{(Color Online) Comparison of the results for the average number of atoms $\bar{N}$ by two methods : direct counting of single fluorescence events (vertical axis) versus fluorescence integration (horizontal axis). Both methods give comparable results, as indicated by the linear fit to the data (dashed line) with a slope of $\sim 1$ (see text). The experiments were performed in two configurations: with a microscopic dipole trap (red diamonds, $w=1~\mu$m) or with a larger dipole trap (blue dots, waist $w=4~\mu$m).}
\label{figure:SingleAtom_MultiAtom}
\end{figure}

This fluorescence integration based method was implemented in two trap configurations that we now describe. As already explained, a significant fraction of the background events come from photons being scattered when the atomic beam interacts with the probe light. To minimize this effect, we decreased the flux of the beam by reducing the temperature of the oven. This is at the expense of the loading rate of the dipole trap: in this first configuration, we could not load more than $\sim10$ atoms in the microscopic dipole trap. In order to calibrate our detection scheme with a larger number of atoms, we implemented a dipole trap with a larger waist $w=4~\mu$m. In this second configuration, the atomic density is smaller and the light-assisted collisions limit the number of atoms to a larger value than in the microscopic dipole trap~\cite{Kuppens2000}, typically $30$. The results of the experiment are shown in Fig.~\ref{figure:SingleAtom_MultiAtom} where we plot the average number of atoms obtained from the counting method versus the average number of atoms obtained by fluorescence integration. The numbers of atoms obtained by the two methods are compatible : a linear fit to the data yields a slope of $1.08$ which is compatible with $1$, as the statistical uncertainty on the slope is $0.05$ and we evaluate the bias uncertainties attached to each method to $4\%$ for the counting method, and $11\%$ for the integrated-fluorescence based method.

\section{Conclusion}
We have demonstrated in this paper a method to reconstruct the distribution of the number of atoms trapped in an optical dipole trap. This method gives access to the distribution of the number of atoms for clouds containing up to a few tens of ultra-cold atoms. It is based on the detection, at the single photon level, of fluorescence events scattered by the atoms when they undergo a random walk in the resonant probe light. So far, we have investigated a regime where the distribution of the number of atoms does not clearly depart from a Poisson law. Our technique is in principle applicable to any distribution and could therefore be useful to investigate blockade effects leading to sub-Poissonian distributions of the number of trapped atoms.

%\end{spacing}
\begin{acknowledgments}
We thank A.M.~Lance and C.~Tuchendler for their contribution at an early stage of the experiment. We acknowledge support from IARPA, the European Union through the Integrated Project AQUTE and the ERC starting grant ARENA, the Institut Francilien des Atomes Froids (IFRAF) and from the PPF `Information quantique' and `Manipulation d'atomes froids  par des lasers de puissance'. A.~Fuhrmanek acknowledges partial support from the DAAD Doktorandenstipendium.
\end{acknowledgments}


\begin{thebibliography}{32}

\bibitem{BlochRMP08} I.~Bloch, J.~Dalibard and W.~Zwerger, Rev. Mod. Phys. {\bf 80} 885 (2008)

\bibitem{Dumke02} R.~Dumke, M.~Volk, T.~M\"uther, F.B.J.~Buchkremer, G.~Birkl and W.~Ertmer, Phys. Rev. Lett.  {\bf 89}, 097903, (2002)

\bibitem{Whitlock09} S.~Whitlock, R.~Gerritsma, T.~Fernholz and R.J.C.~Spreeuw, New J. Phys. {\bf 11} 023021 (2009)

\bibitem{Albiez05} M.~Albiez, R.~Gati, J.~F\"olling, S.~Hunsmann, M.~Cristiani and M.K.~Oberthaler, Phys. Rev. Lett. {\bf 95}, 010402 (2005)

\bibitem{Mandel03} O.~Mandel, M.~Greiner, A.~Widera, T.~Rom, T.W.~H\"ansch and I.~Bloch, Nature {\bf 425}, 937 (2003)

\bibitem{Nelson2007} K.D.~Nelson, X.~Li and D.S.~Weiss, Nat. Phys. \textbf{3}, 556 (2007)

%\bibitem{Weiner99} J. Weiner, V.S. Bagnato, S. Zilio, and P.S. Julienne, Rev. Mod. Phys. {\bf 71}, 1 (1999)
%
%\bibitem{Steane92} A.M. Steane, M. Chowdhury, and C.J. Foot, JOSA B {\bf 12}, 2142 (1992)

%\bibitem{Sokolov09} I.M. Sokolov, M.D. Kupryanova, D.V. Kupryanov and M.D. Havey, Phys. Rev. A {\bf 79} 053405 (2009)

\bibitem{Wineland94} D.J.~Wineland, J.J.~Bollinger, W.M.~Itano and D.J.~Heinzen, Phys. Rev. A {\bf 50}, 67 (1994)

\bibitem{Orzel01} C.~Orzel, A.K.~Tuchman, M.L.~Fenselau, M.~Yasuda and M.A.~Kasevich, Science {\bf 291}, 2386 (2001)

\bibitem{Esteve08} J.~Esteve, C.~Gross, A.~Weller, S.~Giovanazzi and M.K.~Oberthaler, Nature {\bf 455}, 1216 (2008)

\bibitem{Schlosser2001} N.~Schlosser, G.~Reymond, I.~Protsenko and P.~Grangier, Nature (London) {\bf 411}, 1024 (2001)

\bibitem{Forster06} L.~F\"orster, W.~Alt, I.~Dotsenko, M.~Khudaverdyan, D.~Meschede, Y.~Miroshnychenko, S.~Reick and A.~Rauschenbeutel, New. J. Phys. {\bf 8}, 259 (2006)

\bibitem{Chuu2005} C.-S.~Chuu, F.~Schreck, T.P.~Meyrath, J.L.~Hanssen, G.N.~Price and M.G.~Raizen, Phys. Rev. Lett. {\bf 95}, 260403 (2005)

\bibitem{Itah10} A.~Itah, H.~Veksler, O.~Lahav, A.~Blumkin, C.~Moreno, C.~Gordon and J.~Steinhauer, Phys. Rev. Lett. {\bf 104}, 113001 (2010)

\bibitem{Whitlock10} S.~Whitlock, C.F.~Ockeloen and R.J.C.~Spreeuw, Phys. Rev. Lett. {\bf 104}, 120402 (2010)

\bibitem{Bucker09} R.~B\"ucker, A.~Perrin, S.~Manz, T.~Betz, Ch.~Koller, T.~Plisson, J.~Rottmann, T.~Schumm, and J.~Schmiedmayer, New J. Phys. \textbf{11}, 103039 (2009)

%\bibitem{Manz10} S.~Manz, R.~B\"ucker, T.~Betz, Ch.~Koller, S.~Hofferberth, I.E.~Mazets, A.~Imambekov, E.~Demler, A.~Perrin, J.~Schmiedmayer and T.~Schumm, Phys. Rev. A {\bf 81}, 031610(R) (2010)

\bibitem{Sortais07} Y.R.P.~Sortais \textit{et al.}, Phys. Rev. A \textbf{75}, 013406 (2007)

%\bibitem{Sortais07} Y.R.P.~Sortais, H.~Marion, C.~Tuchendler, A.M.~Lance, M.~Lamare, P.~Fournet, C.~Armellin, R.~Mercier, G.~Messin, A.~Browaeys and P.~Grangier, Phys. Rev. A \textbf{75}, 013406 (2007)

\bibitem{Tuchendler08} C.~Tuchendler, A.M.~Lance, A.~Browaeys, Y.R.P.~Sortais and P.~Grangier, Phys. Rev. A \textbf{78}, 033425 (2008)

\bibitem{Fuhrmanek2010} A.~Fuhrmanek, A.M.~Lance, C.~Tuchendler, P.~Grangier, Y.R.P.~Sortais and A.~Browaeys, New. J. Phys. \textbf{12}, 053028 (2010)

\bibitem{Schlosser2002} N.~Schlosser, G.~Reymond and P.~Grangier, Phys. Rev. Lett. {\bf 89}, 023005 (2002)

\bibitem{Kuppens2000} S.J.M.~Kuppens, K.L.~Corwin, K.W.~Miller, T.E.~Chupp and C.E.~Wieman, Phys. Rev. A {\bf 62}, 013406 (2000)

\bibitem{Morice95} O.~Morice, Y.~Castin and J.~Dalibard, Phys. Rev. A {\bf 51}, 3896 (1995)

%~\footnote{The exclusion process is actually subtle in the sense that if the detected peak is surrounded by one or several peaks within a  $13\times 13$ pixels zone, the algorithm excludes these neighboring peaks only if their amplitude is smaller than $11\sigma_{\textrm{CCD}}$. This threshold criterion was determined experimentally by noting that a significant fraction of neighboring peaks with an amplitude below $11\sigma_{\textrm{CCD}}$ appears for large values of the intensifier gain, due to secondary amplification processes. We thus filter out these neighboring peaks, which are artefact events generated by the intensifier. In spite of this filtering, a few artefact events with even larger amplitudes may still be counted, and we evaluate that we may thereby overestimate the number of true events by no more than $1\%$}

%\bibitem{Moore95} F. L. Moore, J. C. Robinson, C. F. Bharucha, Bala Sundaram, and M. G. Raizen, Phys. Rev. Lett. {\bf 75}, 4598 (1995)

%\bibitem{Diener02}  R. B. Diener, B. Wu, M. G. Raizen, and Q. Niu, Phys. Rev. Lett. {\bf 89}, 070401 (2002)

%\bibitem{Bakr09} W.S. Bakr, J.I. Gillen, A. Peng A, S. F\"{o}lling, and M. Greiner, Nature \textbf{462} 74 (2009)


\end{thebibliography}
\end{document}